\documentclass{llncs}
\usepackage[title]{appendix}

\usepackage{amsmath,amssymb,amsfonts,stmaryrd}
\usepackage{graphicx}
\usepackage[usenames,dvipsnames]{color}
\usepackage[utf8]{inputenc}
\usepackage[T1]{fontenc}
\usepackage{booktabs}
\usepackage{xspace}
\usepackage{array}
\usepackage{listings}
\usepackage{algorithm}
\usepackage{algpseudocode}
\usepackage{siunitx}  

\newif\ifcomments%
\commentstrue%

\newcommand{\kd}[0]{k_\text{diff}}
\newcommand{\Xin}[0]{X_\text{in}}
\newcommand{\Xext}[0]{X_\text{ext}}
\newcommand{\Xd}[0]{X_\text{delay}}
\newcommand{\R}[0]{\mathbb{R}}

\begin{document}

\title{On Estimating Derivatives of Input Signals in Biochemistry} 

\author{Mathieu Hemery and François Fages}

\institute{Inria Saclay, Lifeware project-team, Palaiseau, France\\ \email{mathieu.hemery@inria.fr} \email{Francois.Fages@inria.fr}}

\maketitle

\begin{abstract}
  The online estimation of the derivative of an input signal is widespread in control
  theory and engineering.  In the realm of chemical reaction networks (CRN), this raises
  however a number of specific issues on the different ways to achieve it.  A CRN pattern
  for implementing a derivative block has already been proposed for the PID control of
  biochemical processes, and proved correct using Tikhonov's limit theorem.  In this
  paper, we give a detailed mathematical analysis of that CRN, thus clarifying the
  computed quantity and quantifying the error done as a function of the reaction kinetic
  parameters.  In a synthetic biology perspective, we show how this can be used to design error correcting terms
  to compute online functions involving derivatives with CRNs.
  In the systems biology perspective, we give the list of models in BioModels containing (in the sense of subgraph
  epimorphisms) the core derivative CRN, 
  most of which being models of oscillators and control systems in the cell,
  and discuss in detail two such examples: one model of the circadian clock and one model of a bistable switch.
\end{abstract}

\section{Introduction}

Sensing the presence of molecular compounds in a cell compartment is a necessary task of
living cells to maintain themselves in their environment, and achieve high-level functions
as the result of low-level processes of basic biomolecular interactions.  The formalism of
chemical reaction networks (CRN)~\cite{Feinberg77crt} is both a useful abstraction to
describe such complex systems in the perspective of systems biology
\cite{Kitano02science}, and a possible molecular programming language in the perspective
of synthetic biology~\cite{VSK18dna,FLBP17cmsb}.

Sensing the concentration levels of molecular compounds has been well-studied in the
domain of signal transduction networks.  For instance, the ubiquitous CRN structure of
MAPK signaling networks has been shown to provide a way to implement analog-digital
converters in our cells, by transforming a continuous input signal, such as the
concentration of an external hormone activating membrane receptors, into an almost
all-or-nothing output signal according to some threshold value of the input, i.e.~using a
stiff sigmoid as dose-response input-output function~\cite{HF96pnas}.

The analysis of input/output functions fits well with the computational theory of CRNs.  In particular, the
Turing-completeness result shown in~\cite{FLBP17cmsb} for the interpretation by Ordinary Differential Equations (ODE) of CRNs,
possibly restricted to elementary CRNs using mass-action law kinetics and
at most bimolecular reactions, demonstrates the generality of this
approach to biomolecular programming. Furthermore, it comes with an algorithm to automatically generate a finite CRN for
implementing any computable real function.  Such a compiler is implemented
in our CRN modeling software BIOCHAM~\cite{CFS06bi} in several forms, including a
theoretically more limited but practically more interesting framework for robust \emph{online computation}~\cite{HF22cmsb}.

Sensing the derivative of an input molecular concentration is nevertheless beyond the scope
of this computational paradigm since it assumes that the input molecular concentrations are stabilized
at some fixed values which makes no sense for computing the derivative.
Furthermore, it is well-known that the derivative of a computable real function is not
necessarily computable~\cite{Myhill71mmj}.  We must thus content ourselves with
\emph{estimating} the derivative of an input with some error, instead of
\emph{computing} it with arbitrary precision as computability theory requires.

In control theory and engineering, online estimations of input signal derivatives are
used in many places.  Proportional Integral Derivative (PID) controllers adjust a target
variable to some desired value by monitoring three components: the error, that is the
difference between the current value and the target, its
integral over a past time slice, and its current derivative.  The derivative term can
improve the performance of the controller by avoiding overshoots and solving some
problematic cases of instability.

Following early work on the General Purpose Analog Computer (GPAC)~\cite{Shannon41},
the integral terms can be implemented with CRNs using simple
catalytic synthesis reactions such as $A \rightarrow A+B$ for integrating $A$ over time,
indeed $B(T)=\int_O^T A(t) dt$.  Difference terms can be implemented using the
annihilation reaction $A_+ + A_-\rightarrow\emptyset$ which is also used in
\cite{ET89book,OK11iet,FLBP17cmsb} to encode negative values by the difference of two molecular
concentrations, i.e. dual-rail encoding.
This is at the basis of the CRN implementations of, for instance, antithetic PI
controllers presented in~\cite{BGK16cell}.

For the CRN implementation of PID controllers, to the best of our knowledge three different CRN templates have been
proposed to estimate derivative terms. The first one by Chevalier \& al.
\cite{CGNE19cs} is inspired by bacteria's chemotaxis, but relies on strong restrictions upon
the parameters and the structure of the input function making it apparently limited in
scope.
A second one proposed by Alexis \& al.
\cite{alexis2021biomolecular} uses tools from signal theory
to design a derivative circuit with offset coding of negative values
and to provide analytic expressions for its response.
The third one developed by Whitby \& al.~\cite{WCKLT21ieee} is practically similar in its
functioning to the one we study here, differing only on minor implementation details,
and proven correct through Tikhonov's limit theorem.
This result ensures that when
the appropriate kinetic rates tend to infinity, the output is precisely the derivative of
the input.

In this paper, we give a detailed mathematical analysis of that third derivative CRN and quantify the
error done as a function of the reaction kinetic parameters, by providing a first-order
correction term.
We illustrate the precision of this analysis on several examples,
and show how this estimation of the derivative can be actively used with error-correcting terms to compute elementary mathematical
functions online. 
Furthermore, we compare our core derivative CRN to the CRN models in the curated part of \url{BioModels.net} model repository.
For this, we use the theory of subgraph epimorphisms (SEPI)
\cite{GSF10bi,GFMSS14dam} and its implementation in BIOCHAM~\cite{CFS06bi},
to identify the models in BioModels which contain the derivative CRN structure.
We discuss with some details the SEPIs found on two such models:
\verb+biomodels 170+,
one of the smallest eukaryotes circadian clock model~\cite{BWHK04bpj},
and 
\verb+biomodels 318+, a model of the bistable switch at the restriction point of the cell cycle~\cite{yao2008bistable}.

The rest of the article is organized as follow. In Section~\ref{sec:crn}, we provide some
preliminaries on CRNs and their interpretation by ODEs. We present the core differentiation CRN in
Section~\ref{sec:model},  in terms of both of some of its different possible biological
interpretations, and of its mathematical properties.
Section~\ref{sec:analysis} develops the mathematical analysis to bound the error done by that core CRN,
and give in Section~\ref{sec:examples}  some examples to test the validity of our estimation
and the possibility to introduce error-correcting terms.
Section~\ref{sec:biology} is then devoted to the search of that derivative CRN pattern
in BioModels repository and the analysis of those matching in two cases.
Finally, we conclude on the perspectives of our approach to both CRN design at an abstract mathematical level,
and comparison to natural CRNs to help understanding their functions. 

\section{Preliminaries on CRNs} \label{sec:crn}

\subsection{Reactions and Equations}

The CRN formalism allows us to represent the molecular interactions that occur on a finite set
of molecular compounds or species, $\{X_i\}_{i \in 1 \ldots n}$, through a finite set of
formal (bio)chemical reactions, without prejudging their interpretation
in the differential, stochastic, Petri Net and Boolean semantics hierarchy~\cite{FS08tcs}.
Each reaction is a triplet
$(R,P,f)$, also written $ R \xrightarrow{f} P$,
where $R$ and $P$ are multisets of respectively reactant and product species in
$\{X_i\}$, and $f:\R_+^n \mapsto \R_+$ is a kinetic rate function of the reactant species.
A CRN is thus entirely described by the two sets of $n$ species and $m$ reactions:
$\{X_i\},\{R_s \xrightarrow{f_s} P_s\}$.

The differential semantics of a CRN associates positive real valued molecular concentrations,
also noted $X_i$ by abuse of notation,
and the following ODEs which define the time evolution of those concentrations:
\begin{equation}
	\frac{d X_i}{dt} = \sum_{s \in S} (P_s(X_i) - R_s(X_i)) f_s(X),
\end{equation}
where $P_s(X_i)$ (resp. $R_s(X_i)$) denotes the multiplicity (stoichiometry) of $X_i$ in the multiset of products
(resp. reactants) of reaction $s$.


In the case of a mass action law kinetics,
the rate function is a monomial, $f_s = k_s \prod_{x \in R_s} x$,
composed of the product of the concentrations of the reactants by some positive constant $k_s$.
If all reactions have mass action law kinetics, we write the rate constant in place of the rate function $ R \xrightarrow{k} P$,
and the differential semantics of the CRN is defined by a
Polynomial Ordinary Differential Equation (PODE).

From the point of view of the computational theory of CRNs, there is no loss of generality
to restrict ourselves to elementary CRNs composed of at most bimolecular reactions with
mass action law kinetics.  Indeed,~\cite{FLBP17cmsb} shows that any computable real
functions (in the sense of computable analysis, i.e. with arbitrary finite precision by a
Turing machine), can be computed by such a CRN, using the dual-rail encoding of real
values by the difference of molecular concentrations, $x=X_+-X_-$. While our compiler
ensures that the quantity $X_+-X_-$ behaves properly, it is also important to degrade
both of them with an annihilation reaction, $X_+ + X_- \xrightarrow{fast} \emptyset$,
to avoid a spurious increase of their concentration.
Those annihilation reactions are supposed to be faster than the other reactions of the CRN.

\begin{example}
   The first example given in \cite{FLBP17cmsb} showed the compilation of the cosine function of
   time, $y=cos(t)$ in the following CRN:
  \begin{equation}
	  \label{CRN:compiled_cosine}
	   \begin{aligned}
          A_p  \rightarrow A_p+y_p &\quad\quad\quad
          A_m  \rightarrow A_m+y_m &  \quad\hspace{1cm} & A_m(0)=0,\ A_p(0)=0\\
          y_m  \rightarrow A_p+y_m &\quad\quad\quad
          y_p  \rightarrow A_m+y_p & \quad\hspace{1cm}  & y_m(0)=0,\ y_p(0)=1\\
          y_m+y_p  \xrightarrow{fast} \emptyset &\quad\quad\quad
          A_m+A_p  \xrightarrow{fast} \emptyset&&
      \end{aligned}
   \end{equation}

  The last two reactions are necessary to avoid an exponential increase of the species concentration. 
  The associated PODE is:
  \begin{equation}
	\begin{aligned}
d(A_m)/dt &= y_p-fast*A_m*A_p & \hspace{1cm} & A_m(0)&=0\\
d(A_p)/dt &= y_m-fast*A_m*A_p & \hspace{1cm}  & A_p(0)&=0\\
d(y_m)/dt &= A_m-fast*y_m*y_p & \hspace{1cm}  & y_m(0)&=0\\
d(y_p)/dt &= A_p-fast*y_m*y_p & \hspace{1cm}  & y_p(0)&=1
        \end{aligned}
        \end{equation}

\end{example}

\subsection{CRN Computational Frameworks}

The notions of CRN computation proposed in~\cite{FLBP17cmsb} and~\cite{HF22cmsb}
for computing input/ouput functions, do not provide however
a suitable framework for computing derivative functions.
Both rely on a computation at
the limit, meaning that the output converges to the result of the computation whenever the CRN is either properly
initialized~\cite{FLBP17cmsb}, or the inputs are stable for a sufficient period of
time~\cite{HF22cmsb}. To compute a derivative, we cannot ask that the input stay fixed for
any period of time as this would imply a null derivative. We want the output to follow
«~at run time~» the derivative of the input.

Our question is thus as follows. Given an input species $X$ following a time course
imposed by the environment $X(t)$, is it possible to perform an online computation such
that we can approximate the derivative $\frac{dX}{dt}$ on the concentration of $2$ output
species using a dual-rail encoding?

The idea is to approximate the left derivative by getting back to its very mathematical
definition:
\begin{equation}\label{eq:definition}
	\frac{dX}{dt}(t) = \lim_{\epsilon \rightarrow 0^+} \frac{X(t)-X(t-\epsilon)}{\epsilon},
\end{equation}
but how can we measure $X(t-\epsilon)$?

\section{Differentiation CRN} \label{sec:model}

\subsection{Biological intuition using a membrane}

\begin{figure}
	\centering
	\includegraphics[width=0.6\textwidth]{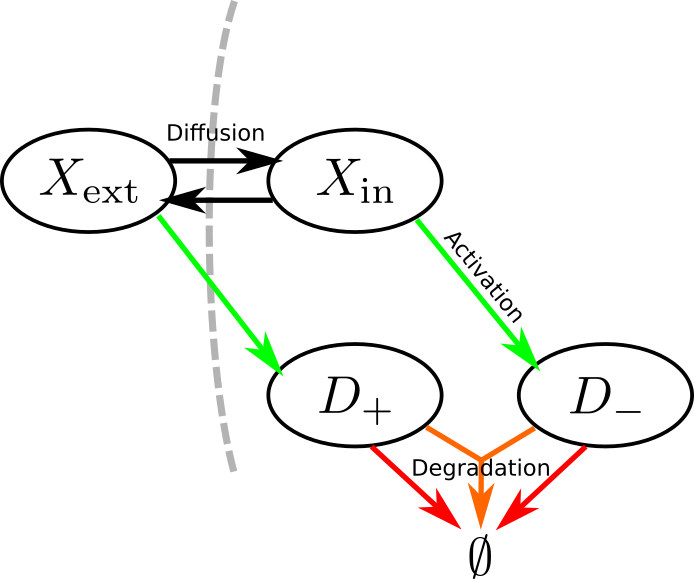}
	\caption{Hypergraph representation of the core differentiation CRN composed
	of one input, two outputs and one intermediate species. The input $\Xext$ is
   outside of the membrane and thus present in so large quantity that its concentration is not
   modified by the dynamics, once it crossed the membrane it is labelled as $\Xin$. Each species
   $\Xext$ (resp. $\Xin$) activates the synthesis of its part of the input: $D_+$ (resp.
   $D_-$). Finally a fast annihilation reaction eliminates both $D_+$ and $D_-$ so that only the
   highest of the two remains present, depending on the sign of the derivative of $\Xext$.}
	\label{fig:model}
\end{figure}

One biological intuition we may have to measure a value in a previous time is to use a
membrane with a fast diffusive constant. Indeed, if we suppose that the input is the
outside species, the inside species equilibrates to follow the concentration of the
outside one (the input) but also suffers a lag due to the diffusion. Building upon
this simple trick leads to the CRN presented in Fig.~\ref{fig:model}.
As the derivative may be positive or negative, a dual-rail encoding is used for the derivative.
This CRN is
mainly equivalent to the derivative block proposed in~\cite{WCKLT21ieee} apart from the
fact that we suppose (for the sake of clarity) that the input stay positive and no dual-rail encoding is used for it.
In the case of a
dual-rail encoded input, the two species need to have the same permeability through
the membrane, otherwise the delay is not the same for the positive and
negative parts.

The delay is thus introduced through a membrane
under the assumption that the outside concentration is imposed by the environment. This
conveniently explains why the kinetic rates are the same for the two monomials in the
derivative of $\Xin$, but this is not mandatory.
Indeed two other settings can be used to construct such a CRN without relying on a
membrane.  We could use a phosphorylation and a dephosphorylation reactions where $\Xin$
would be the phosphorylated species.  Or we could, as in~\cite{WCKLT21ieee}, rely on a
catalytic production of $\Xin$ by $\Xext$ and a degradation reaction of $\Xin$. A drawback
of these two other implementations is that they need to be tuned to minimize the
difference between the rates of the two monomials in the derivative of $\Xin$. Otherwise
a proportional constant is introduced between $\Xext$ and $\Xin$, and needs to be
corrected by adjusting the production rates of $D_+$ and $D_-$.

However, the membrane implementation also has its own drawback as it requires the reaction
$\Xext \rightarrow \Xext + D_+$ to occur through the membrane. We may think of a membrane
protein $M$ that mediates this reaction ($\Xext + M \rightarrow \Xext + M + D_+$). Then, since
its concentration is constant, it can simply be wrap up in the kinetic constant of the reaction.
Which of this three implementations should be chosen may depend on the exact details of
the system to be build.

\subsection{Core differentiation CRN}

Our core differentiation CRN schematized in Fig.~\ref{fig:model}
is more precisely composed of the following 7 reactions:
\begin{equation}\label{eq:CRN}
\begin{aligned}
	\Xext &\xrightarrow{k_\text{diff}} \Xin &
	\Xin &\xrightarrow{k_\text{diff}} \Xext \\
	\Xext &\xrightarrow{k.k_\text{diff}} \Xext + D_+ &
	\Xin &\xrightarrow{k.k_\text{diff}} \Xin + D_- \\
	D_+ &\xrightarrow{k} \emptyset &
	D_- &\xrightarrow{k} \emptyset \\
	D_+ + D_- &\xrightarrow{\text{fast}} \emptyset
\end{aligned}
\end{equation}
The diffusion through the membrane is symmetrical with a constant $\kd$ and both activations
should have the same constant product $k.\kd$ while the degradation of the outputs should have a
rate $k$.
We make the assumption that the outside species $\Xext$ is present in large
quantity so that its concentration is not affected by the dynamics of the CRN.
Under this assumption, the differential
semantics is then the same as the one of the differentiation CRN
proposed in~\cite{WCKLT21ieee}:
\begin{equation}\label{eq:PODE}
\begin{aligned}
	\frac{d\Xin}{dt} &= k_\text{diff} (\Xext - \Xin) \\
	\frac{dD_+}{dt} &= k k_\text{diff} \Xext - k D_+ - \text{fast} D_+ D_- \\
	\frac{dD_-}{dt} &= k k_\text{diff} \Xin - k D_- - \text{fast} D_+ D_-
\end{aligned}
\end{equation}

The derivative is encoded as $D = D_+ - D_-$ and hence obeys the equation
(using the two last lines of the previous equation):
\begin{equation}\label{eq:derivative}
\begin{aligned}
	\frac{dD}{dt} &= \frac{dD_+}{dt} - \frac{dD_-}{dt}  \\
					  &= k k_\text{diff} (\Xext - \Xin) - k (D_+ - D_-) \\
	\frac{dD}{dt} &= k \left( \frac{\Xext - \Xin}{\frac{1}{\kd}} - D \right)
\end{aligned}
\end{equation}

In the next section, we prove that $\Xin$ is equal to $\Xext$ with a delay $\epsilon$,
hence giving us our second time point $X(t-\epsilon)$,
up to the first order in
$\epsilon = \frac{1}{\kd}$.
The fractional part of the last equation is thus precisely an
estimate of the derivative of $\Xext$ as defined in Eq.~\ref{eq:definition}, with a
finite value for $\epsilon$.

It is also worth remarking that such derivative circuits can in principle be connected to
compute higher-order derivatives, with a dual-rail encoded input. It is
well known that such estimations of higher-order derivatives can be very sensitive to
noise and error, and are thus not reliable for precise computation but may be good enough
for biological purposes.  We will see a biological example of this kind in
Section~\ref{ss:clock} on a simple model of the circadian clock.

\section{Mathematical analysis of the quality of the estimation} \label{sec:analysis}

Our first goal is to determine precisely the relation between $\Xin$ and $\Xext$ when the
later is enforced by the environment. Using the first line of Eq.~\ref{eq:PODE}, we
obtain by symbolic integration:
\begin{equation}
	\Xin(t) = \kd \int_0^\infty \exp(-\kd s) \Xext (t-s) ds,
\end{equation}
where we can see that $\Xin$ is the convolution of $\Xext$ with a decreasing exponential.
This convolution is not without reminding the notion of \emph{evaluation} in the theory of
distribution and has important properties of regularisation of the input function. In
particular, whatever the input function is, this ensures that the internal representation
is continuous and differentiable.

The interesting limit for us is when $\kd \rightarrow \infty$, that is when $\epsilon =
\frac{1}{\kd} \rightarrow 0$. In this case, the exponential is neglectable except in a
neighbourhood of the current time and supposing that $\Xext$ is infinitely
differentiable\footnote{We also explore in Figures~\ref{fig:delay}\textbf{D} and
  \ref{fig:derivative}\textbf{C} what a non analyticity of $\Xext$ imply for our model.},
we obtain by Taylor expansion:
\begin{equation}
\begin{aligned}
	\Xin(t) &= \int_0^\infty \kd \exp(-\kd s) \sum_{n=0}^\infty \frac{(-s)^n}{n!}
	\Xext^{(n)}(t) ds \\
	 &= \sum_{n=0}^\infty \frac{\kd}{n!} \Xext^{(n)}(t) \int_0^\infty (-s)^n \exp(-\kd s)
	 ds
\end{aligned}
\end{equation}

The integral may be evaluated separately using integration by parts and recursion:
\begin{equation}
\begin{aligned}
	I_n &= \int_0^\infty (-s)^n \exp(-\kd s) ds = -n \epsilon I_{n-1} \\
	 &= (-1)^n (\epsilon)^{n+1} n!
\end{aligned}
\end{equation}

We thus have:
\begin{equation}
	\label{eq:delay_approx}
\begin{aligned}
	\Xin(t) &= \sum_{n=0}^\infty \frac{\kd}{n!} \Xext^{(n)}(t) (-1)^n n! \epsilon^{n+1} \\
   &= \sum_n (-\epsilon)^n \Xext^{(n)}(t) \\
	&= \Xext(t) - \epsilon \Xext'(t) + \epsilon^2 \Xext''(t) + \ldots \\
	\Xin(t) &= \Xext \left( t - \epsilon \right) + o(\epsilon^2).
\end{aligned}
\end{equation}

Using Taylor expansion once again in the last equation somehow formalizes
our intuition: the concentration of the internal species $\Xin$ follows the time course
of the external one with a delay equal to the inverse of the diffusive constant $\kd$.
This validates our formulation of the derivative.

Now, it is sufficient to remark that Eq.~\ref{eq:derivative} has exactly the same
form as the first line of Eq.~\ref{eq:PODE} that we just study in length. Just
replace $\Xext$ by the estimation of the left derivative, $\Xext$ by the output $D$ and
the rate constant $k$ instead of $\kd$. The delay approximation is thus also possible in
this step and, introducing the delay $\tau = \frac{1}{k}$, we immediately obtain a precise
expression for $D$:

\begin{equation}
   D(t) = \frac{\Xext(t-\tau) - \Xext(t-\epsilon-\tau)}{\epsilon}+o(\epsilon)+o(\tau^2).
\end{equation}

We can see this as the secant approximation of the derivative of $\Xext$ with a step
size $\epsilon$ and a delay $\tau$. Moreover we also know that the residual error on this
expression are of first order in $\epsilon$ and second order in $\tau$.

It is well known in the field of numerical computation that the secant method provides a
rather poor approximation, but it has the benefit to be the simplest one, and thus
gives here a small size derivative circuit.
In the hope of improving the precision, one could implement higher-order methods using
several "membranes" to access the value of the function on several time points
before performing the adapted computation.
Such complexation would however also increase the delay
between the input and output function.

\section{Validation on simple examples} \label{sec:examples}

\subsection{Verification of the delay-approximation}

In this first subsection, we want to validate the approximation expressed by
Eq.~\ref{eq:delay_approx}. For this, we focus on the diffusion part of our CRN:
$\Xext \leftrightarrow \Xin$. We make numerical simulation for $2$ different values of
$\epsilon$ and $2$ different input functions: a sine wave and an absolute value signals.
The second allowing us to see how well the delay approximation works in presence of
non analyticity.

\begin{figure}
	\centering
	\includegraphics[width=\textwidth]{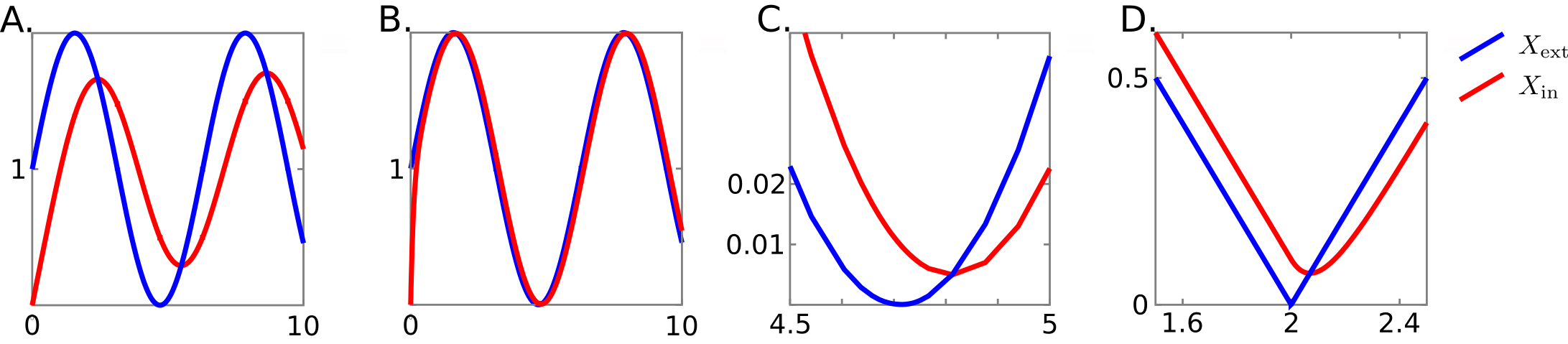}
	\caption{Behaviour of the simple diffusion model when fed with different input
	functions. We have $\kd = 1$ in panel \textbf{A} and $\kd = 10$ in the three others.
	The input function of the three first panels is an offset sine wave: $\Xext(t) =
	1+\sin(t)$.  Panel \textbf{C} is a focus on part of the panel \textbf{B} Panel
	\textbf{D} present as input a shifted absolute value: $\Xext(t) = |t-2|$ as a study
	case of non-differentiable function. See the main text for the discussion.}
	\label{fig:delay}
\end{figure}

Fig.~\ref{fig:delay} shows the response of $\Xin$ in that different condition. In
panel \textbf{A}, the kinetic constant is very low so we expect our approximation to
fail. Indeed, one can see that in addition to having an important delay, the output is strongly
smoothed, this tends to average the variation of the input, bringing back $\Xin$ to the
average value of the input. In panel \textbf{B} the diffusion constant is
increased by a factor $10$. The delay approximation is now very good and we only expect an
error of order $\epsilon^2 = 10^{-2}$ which can be checked with good accuracy on panel
\textbf{C}. Panel \textbf{D} shows a case of a
non-differentiable function in which an error of order $\epsilon = 0.1$ is visible shortly
after the discontinuity and vanishes in a similar timescale.

\subsection{Approximation of the derivative}

\begin{figure}
	\centering
	\includegraphics[width=\textwidth]{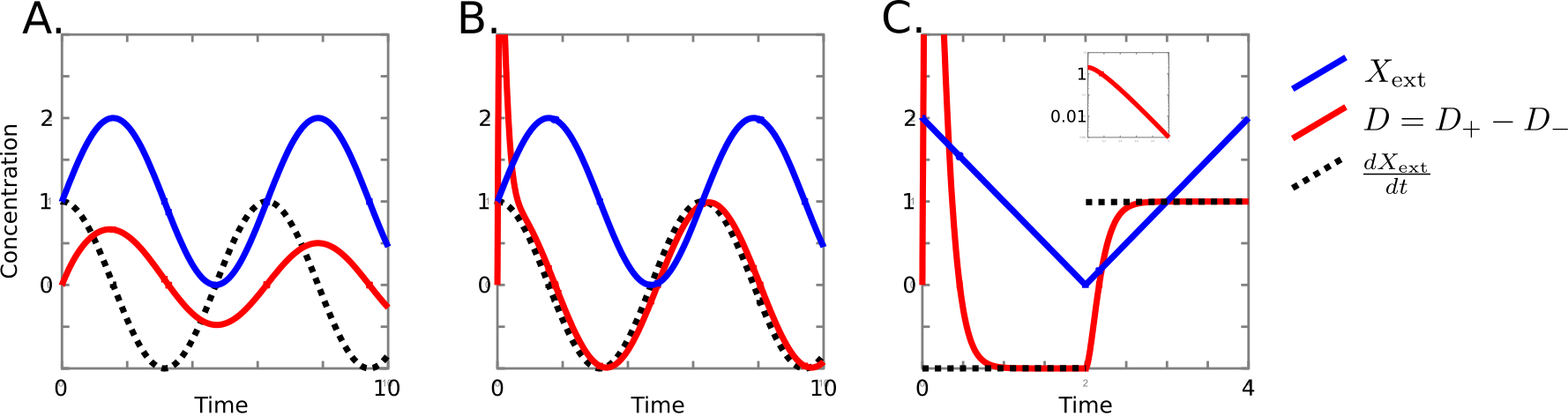}
	\caption{Simulation of the derivative circuit for $\kd = k = 1$ (panel \textbf{A}) 
	and $\kd = k = 10$ (panels \textbf{B} and \textbf{C})
	when fed with two different input signal: a sine wave (panels \textbf{A} and
	\textbf{B}) and an offset absolute value (panel \textbf{C}).
	In \textbf{A} the characteristic time of the diffusion is too large ($\epsilon = 1$) so
	that the delay approximation fail.
	In \textbf{B} $\epsilon = {10}^{-1}$, the approximation is sounded and the output $D = D_+ -
	D_-$ correctly follows a cosine (in dark) with a delay of $\tau = {10}^{-1}$.
	In \textbf{C} the
	singularity makes the derivative harder to compute. As for the diffusion, we see
	a typical decreasing exponential toward its correct value as shown in the inset that
   depict the absolute difference between $D$ and its correct value ($1$ in this case) in
   logarithmic scale for the time between $2$ and $3$. (That is precisely the same time as
   the main figure where the inset is placed.)} 
	\label{fig:derivative}
\end{figure}

Let us now check the behaviour of the derivative circuit. On Fig.~\ref{fig:derivative},
we can see the response of our derivative circuit for a sine wave and an absolute value
input functions. In panels \textbf{A} and \textbf{B} we see that when the first and second order
derivatives of the input are smaller than the kinetic reaction rates, the delay
approximation gives a very good picture of the response. On a complementary point of view,
the panel \textbf{C} shows that in front of singularity, the system adapts after an
exponential transient phase with a characteristic time $\tau = \frac{1}{k}$.

\subsection{Using signal derivatives for online computations}

Our main motivation for analyzing the differentiation CRN is to
compute a function $f$ of some unknown input signal, $\Xext(t)$, online.
that is, given a function $f$, compute a function $f(\Xext(t))$ 
Yet the differentiation CRN only
allows us to approximate the derivative of the input signal.
The idea is thus to implement the PODE:
\begin{equation}
	\frac{dY}{dt} = f'(\Xext(t)) \frac{d \Xext}{dt},\hspace{1cm} Y(0)=f(X(0)
\end{equation}
and provide the
result online on a set of internal species $Y(t) = Y_+ - Y_-$.
This necessitates to compute the function $f'$ and estimate the derivative of the input.
Using the formalism developed in \cite{HFS20cmsb,HFS21cmsb} we know that there
exist an elmentary CRN (i.e.~quadratic PODE) computing $f'(\Xext)$ for any elementary function $f$ and we just
have shown that $\frac{d \Xext}{dt}$ can be approximated by the differentiation CRN.
Therefore, in principle, any elementary function of input signals can be approximated online by a CRN.

As a toy example, let us consider the square function, $\frac{d Y}{dt} = 2 \Xext (D_+ - D_-)$,
and as input, a sine wave offset to stay positive : $\Xext(t) = 1 + \sin(t)$.

The CRN generated by BIOCHAM according to these principles, to compute the square of the input online is:
\begin{equation}
	\begin{aligned}
		\Xext &\xrightarrow{\kd} \Xin, &\quad \Xin &\xrightarrow{\kd} \Xext, \\
		\Xext &\xrightarrow{\kd k} \Xext + D_+ &\quad D_+ &\xrightarrow{k} \emptyset \\ 
		\Xin &\xrightarrow{\kd k} \Xin + D_- &\quad D_- &\xrightarrow{k} \emptyset \\ 
		\Xin + D_+ &\xrightarrow{2} \Xin + D_+ + Y_+ &\quad
		\Xin + D_- &\xrightarrow{2} \Xin + D_- + Y_- \\
		D_+ + D_- &\xrightarrow{\text{fast}} \emptyset &\quad
		Y_+ + Y_- &\xrightarrow{\text{fast}} \emptyset \\
	\end{aligned}
\end{equation}
The first three lines implement the derivative circuit, the fourth line implements
the derivative of $Y$ and the last line provides the dual-rail encoding.

\begin{figure}
	\centering
	\includegraphics[width=\textwidth]{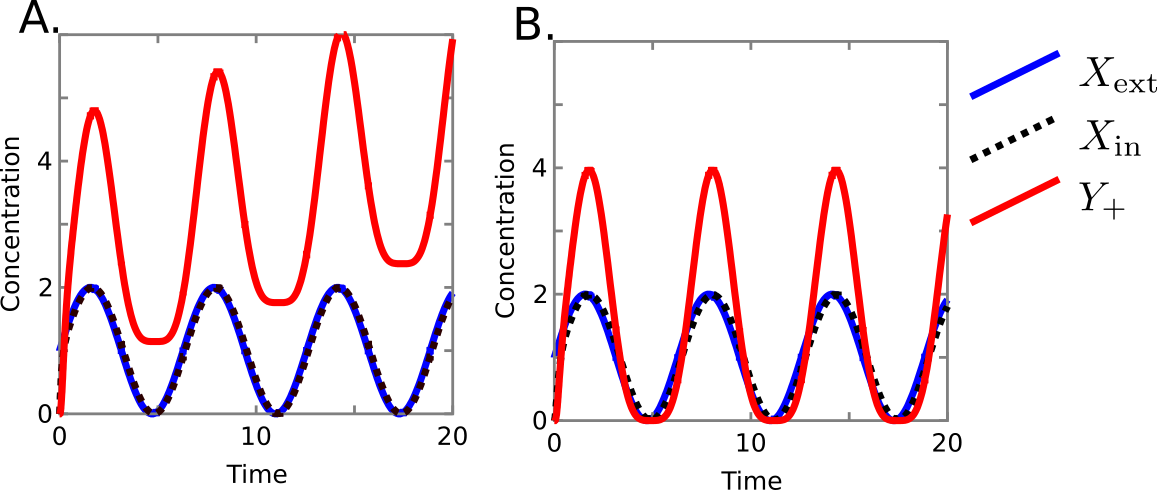}
	\caption{Square computation CRN when fed with an offset sine wave $\Xext(t) =
   1+\sin(t)$. The parameters are: $\kd = k = 10, \text{fast}={10}^6$. Panel \textbf{A}
   shows a simulation of the naive CRN while panel \textbf{B} shows the corrected one
   where the derivative is integrated using a delayed input which eliminates the drift
   presented in the first panel. } 
	\label{fig:computation}
\end{figure}

The numerical simulation of this CRN is depicted in Fig.~\ref{fig:computation}\textbf{A}
One can see that while it effectively computes the square of the input, it also suffers
from a strong drift.  To verify if this drift comes from the delay between the input and
the output, we can compute analytically the output of our network with our approximation of
derivative with a delay (see the full computation in Appendix).
\begin{equation}
   \label{eq:offset}
	\begin{aligned}
		y(t) &= \int 2 x(s) x'(s-\tau) ds \\
		  &\simeq \left(1+\sin(t)\right)^2 + \tau t.
	\end{aligned}
\end{equation}
This is precisely the behaviour that can be seen on the time course of Fig.~\ref{fig:computation}
\textbf{A}. After the integration of $20$ time units, the offset is of order $2$ which is
exactly what is predicted for a delay $\tau = \frac{1}{k} = 0.1$.  Therefore,
while it is always possible to get rid of such errors by increasing $\kd$, 
the identification of the cause of the drift, gives us a potentially simpler path to eliminate it:
using a representation of the input that is itself delayed: $\Xin \leftrightarrow
X_\text{delay}$, and use this delayed signal as the catalyst for the production of $Y_+$
and $Y_-$ in the place of $\Xin$. This leads to the CRN given in Appendix
(Eq.~\ref{CRN:corrected}) for which numerical integration shows in
Fig.~\ref{fig:computation}\textbf{B} that we indeed have get rid of the drift, or said
otherwise, the correct implementation for online computation is given by:
\begin{equation}
   \frac{dY}{dt} = f'(\Xext(t-\tau)) \frac{d\Xext}{dt}(t-\tau),
\end{equation}
where the delays has to be equal for the two pieces of the derivative.


\section{Biological examples}
\label{sec:biology}

\subsection{BioModels repository}\label{sec:biomodels}

\begin{figure}
   \centering
   \includegraphics[width=0.6\textwidth]{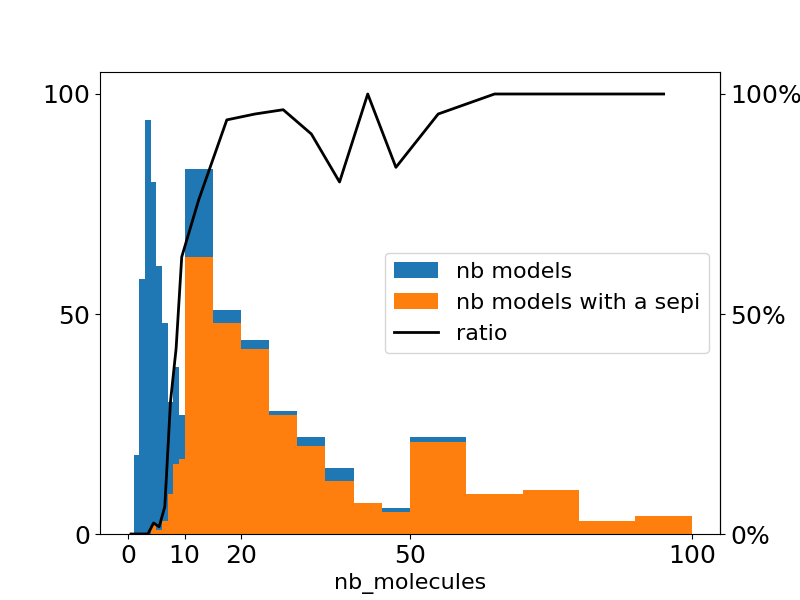}
   \caption{Number of models in the curated part of BioModels per number of species,
     given with the number of models having a SEPI reduction to the differentiation CRN and ratio (black curve) between these two quantities.
   }
   \label{fig:cutoff}
\end{figure}

To explore the possibility that natural biochemical systems already implement a form or another of
the core differentiation CRN, one can try to scan the CRN models of the BioModels
repository~\cite{NBBCDDLSSSSH06nar}.
This can be automated with the general graph matching notion of Subgraph EPImorphism (SEPI)
introduced in~\cite{GSF10bi,GFMSS14dam} to compare CRN models
and identify model reduction relationships based on their graph structures.
SEPI generalizes the classical notion of subgraph isomorphism by introducing an operation of node merging in addition to node deletion.
Considering two bipartite graphs of species and reactions, there exists a SEPI from $G_A$
to $G_B$ if there exists a sequence of mergings\footnote{A species (resp. reaction) node can
only be merged with another species (resp. reaction) node and the resulting node inherits of all the
incoming and outcoming edges of the two nodes.} and deletions of nodes in $G_A$
such that the resulting graph is isomorphic to $G_B$.

More precisely, we used the SEPI detection algorithm of BIOCHAM to scan the
curated models in Biomodels (after automatic rewriting with well-formed reactions \cite{FGS15tcs})
and check the existence of a SEPI from each model graph to the differentiation CRN graph.
Fig.~\ref{fig:cutoff} shows that our small differentiation CRN with 4 species is frequently found in large models.
It is thus reasonable to restrict to models with no more than 10 species.
Table~\ref{table} lists the models with no more than 10 species in the 700 first models of BioModels
that contain our differentiation CRN.
The predominance of models exhibiting oscillatory dynamics, and in particular circadian clock models is striking.

\begin{table}
	\begin{center}
      \resizebox{!}{0.45\textheight}{%
		\begin{tabular}{|c||c|c|l|}
			\hline
			Model ID & \# Species & \# reactions & Topic \\
			\hline
0021 & 10 & 30 & Circadian clock \\
0022 & 10 & 34 & Circadian clock \\
0034 &  9 & 22 & Circadian clock \\
0035 &  9 & 15 & Circadian clock \\
0041 & 10 & 17 & Creatine kinase \\
0065 &  8 & 16 & Operon lactose \\
0067 &  7 & 16 & Circadian clock \\
0075 & 10 & 13 & Phosphoinositide turnover \\
0084 &  8 & 16 & ERK Cascade \\ 
0107 &  9 & 23 & Cell cycle \\
0108 &  9 & 18 & Superoxide dismutase overexpression \\
0170 &  7 & 17 & Circadian clock \\ 
0171 & 10 & 27 & Circadian clock \\ 
0179 & 7 & 17 & Cellular memory \\ 
0185 &  8 & 20 & Circadian clock \\
0206 &  9 & 22 & Circadian clock \\
0206 &  8 & 15 & Glycolytic oscillations\\
0216 &  5 & 17 & Circadian clock \\
0228 &  9 & 22 & Cell cycle \\
0229 &  7 & 28 & Circadian clock \\
0240 &  6 & 14 & DegU transcriptional regulator \\
0257 &  8 & 19 & Self-maintaining Metabolism \\ 
0262 &  9 & 14 & AkT Signalling \\ 
0263 &  9 & 14 & AkT Signalling \\ 
0269 &  9 & 22 & Hormonal crosstalk in plant \\
0318 &  7 & 17 & Bistable switch \\
0355 &  9 & 17 & Calcium signalling \\
0359 &  9 & 15 & Tissue factor pathway inhibitor \\
0360 &  9 & 15 & Tissue factor pathway inhibitor \\
0495 &  8 & 18 & Phospholipid synthetic pathways \\
0530 & 10 & 17 & Cooperative gene regulation \\
0539 &  6 & 11 & Mixed feedback loop \\
0563 & 10 & 17 & Plant-microbe interaction \\
0586 & 10 & 23 & Genetic oscillatory network \\
0587 & 10 & 23 & Genetic oscillatory network \\
0590 &  9 & 40 & Biosynthesis of pyrimidines \\
0615 &  4 & 34 & Aggregation kinetics in Parkinson's Disease \\
0616 &  4 & 20 & Resolution of inflammation \\
0619 & 10 & 13 & Basic model of Acetaminophen \\
0622 & 10 & 19 & Ubiquitination oscillatory dynamics \\
0632 &  8 & 14 & Cell fate decision \\
0665 &  7 & 13 & Interleukin-2 dynamics \\
0696 &  9 & 23 & Incoherent type 1 feed-forward loop \\
			\hline
      \end{tabular}}
	\end{center}
        \caption{List of models having a SEPI reduction to the differentiation CRN, given
		  with model ID, number of species, number of reactions, and process modeled,
          among the first $700$ models of the curated part of Biomodels with no more than $10$ species.}\label{table}
\end{table}

\subsection{Circadian clock}\label{ss:clock}

\begin{figure}
	\centering
	\includegraphics[width=\textwidth]{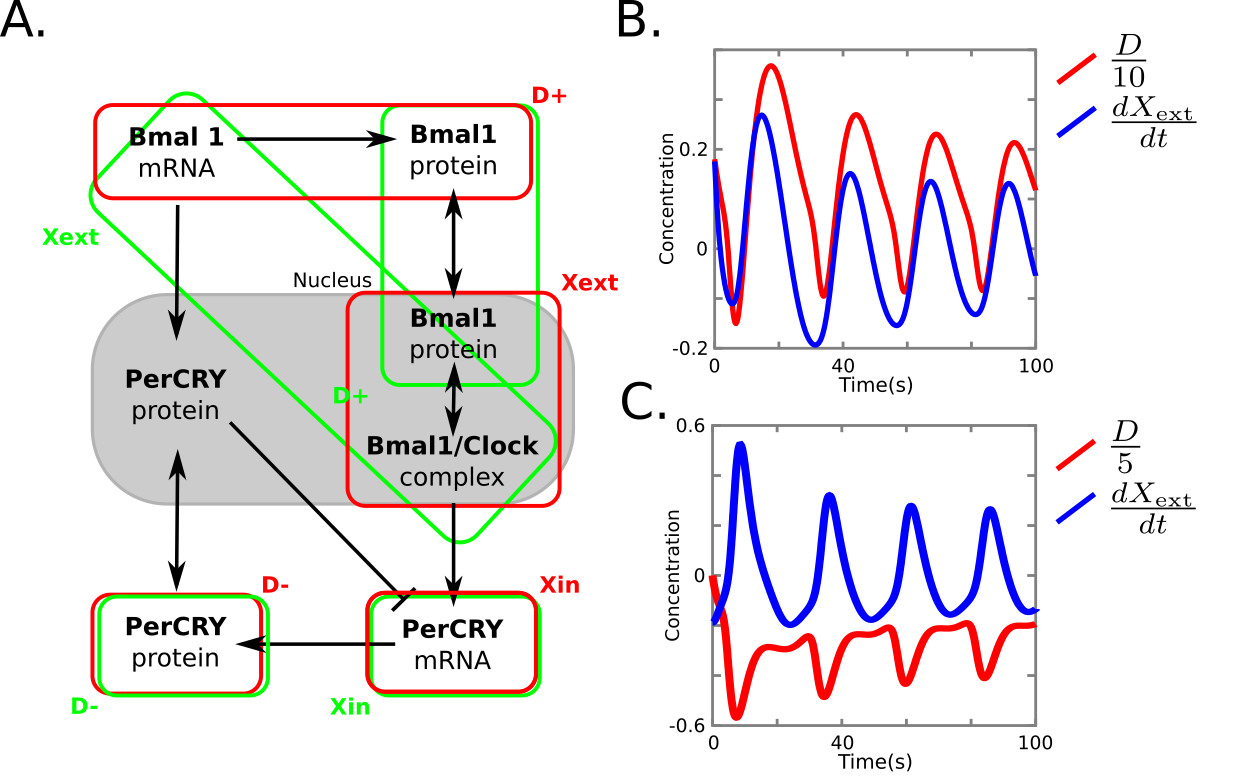}
	\caption{Circadian clock model proposed by Becker-Weimann \& al.
		\textbf{A} Graph of the model, the grey block indicates the nucleus
		while the other species are present in the cytoplasm. The first
		SEPI found by BIOCHAM is displayed in red.
                A second SEPI is displayed in green where the Per/CRY
		complex in the nucleus is mapped to the input of our differentiation CRN.
	\textbf{B} Validation of the first (red) SEPI: we display the derivative of the input
	species here considered as the sum of its part: $\Xext =
	\text{Bmal1}^\text{nucleus}_\text{protein} +
	\text{Bmal1/Clock}^\text{nucleus}_\text{protein}$ and similarly for the output: $D =
	D_+ - D_- = \text{Bmal1}^\text{cytoplasm}_\text{mRNA} + \text{Bmal1}^\text{cytoplasm}_\text{protein}	- \text{Per/CRY}^\text{cytoplasm}_\text{protein}$
	\textbf{C} Validation of the second (green) SEPI. As for the previous panel, species
	that are mapped together are simply summed for this validation.
	The qualitative matching of the two quantities in these two graphs are a good indication that
	these SEPI are meaningful.}
	\label{fig:oscillator}
\end{figure}

Model \verb+biomodels 170+ of the eukaryotes circadian clock proposed by Becker-Weimann \& al.
\cite{BWHK04bpj} is among the smallest models of the circadian clock displaying a SEPI reduction
toward our differentiation CRN. Its influence graph is depicted in
Fig.~\ref{fig:oscillator}\textbf{A}, we also display in red the first SEPI found by BIOCHAM,
and in green a second one obtain by enforcing the mapping from the PER/Cry
species inside the nucleus to the input of the differentiation CRN.
Interestingly, this model has the nucleus membrane separating the species 
mapped to $\Xext$ and the one mapped to $\Xin$ in the second SEPI. 
The oscillatory behavior of this model is shown in panel \textbf{B}.

Now, thinking at the mathematical insight that this relation provides, it is quite natural
for a CRN implementing an oscillator to evaluate its own derivative on the fly.
Actually, when looking at the natural symmetry of the model, we are inclined to think that this
CRN may actually be two interlocked CRNs of the derivative circuit, both computing
the derivative of the output of the other, as if a second order derivative circuit was
closed on itself.
This is something we could easily check by imposing restrictions on the
SEPI mapping. Enforcing the nucleus PerCRY protein to be mapped on $\Xext$ gives us the
SEPI shown in green in Fig.~\ref{fig:oscillator}\textbf{A}.
To validate the preservation of the function of the derivative CRN given by this SEPI,
we can verify that the quantities defined by summing the species
that are mapped together are effectively linked by the desired derivative relation. As
can be seen in Fig.~\ref{fig:oscillator}\textbf{B}, the agreement is striking.
One can even note that the delay of the chemical derivative is the one predicted by our theory.

The case of Fig.~\ref{fig:oscillator}\textbf{C} is more complex as this part of the model seems to compute the opposite of the derivative.
It is however worth noting that there is absolutely no degree
of freedom in our choice of the species used in Fig.~\ref{fig:oscillator}\textbf{B} and
\textbf{C} that are entirely constrained by the SEPI given by BIOCHAM.
Taking both SEPI together we
see that $\text{Bmal1}^\text{nucleus}_\text{protein}$ and
$\text{Bmal1}^\text{cytoplasm}_\text{mRNA}$ play symmetrical roles, being the input and
derivative of the two displayed SEPI. Given that the second SEPI introduces a negative
sign, we may see this as:
\begin{equation}
	\begin{aligned}
		\text{Bmal1}^\text{cytoplasm}_\text{mRNA} &= \frac{d}{dt}
		\text{Bmal1}^\text{nucleus}_\text{protein} \\
		\text{Bmal1}^\text{nucleus}_\text{protein} &= -\frac{d}{dt} \text{Bmal1}^\text{cytoplasm}_\text{mRNA}
	\end{aligned}
\end{equation}
The solution of this well known equation are the sine and cosine functions, and this
perfectly fits the oscillatory behaviour of this CRN. To confirm this hypothesis, we check
for the presence of a SEPI from the clock model to the compiled cosine CRN presented in
Eq.~\ref{CRN:compiled_cosine} which is effectively the case.
On the other hand, there
is no SEPI relation between the compiled cosine and the derivative circuit.

\subsection{Bistable switch}

\begin{figure}
   \centering
   \includegraphics[width=\textwidth]{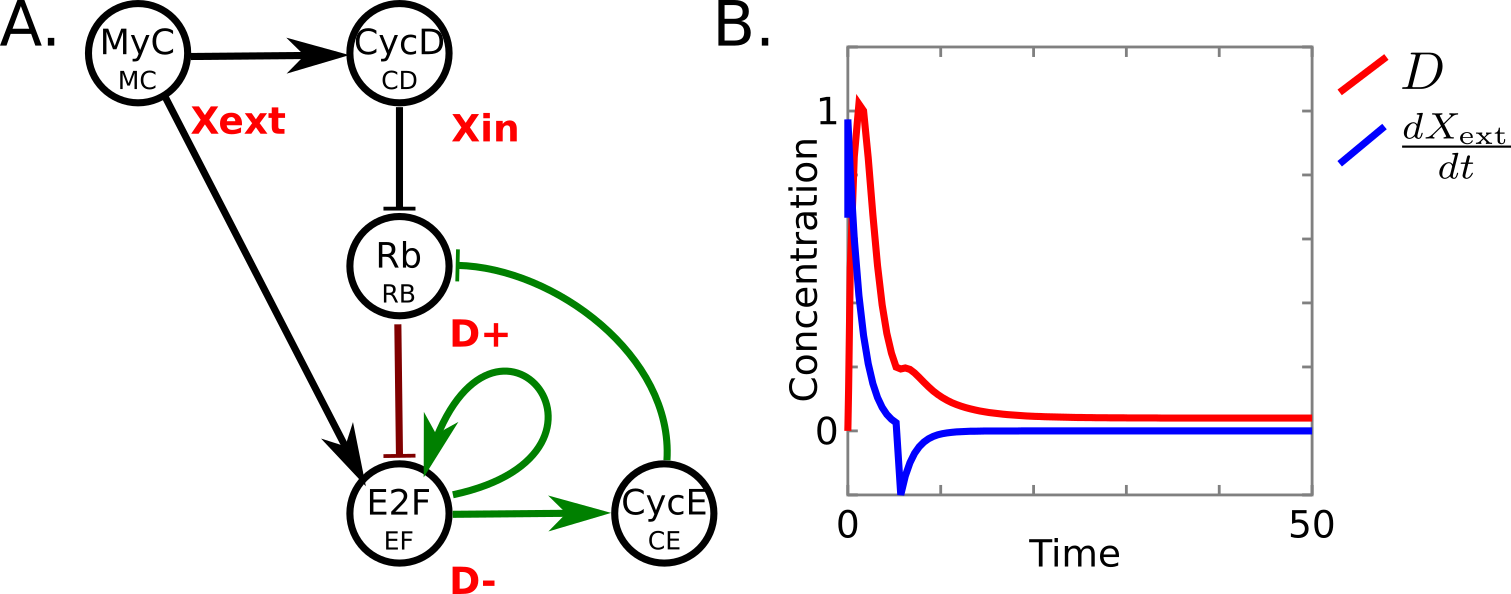}
   \caption{Analysis of the bistable switch. \textbf{A} Schematic representation of the
   model and its SEPI, the smaller fonts for the species corresponds to the names used in
	the model provided online. Two other species are also present in the model (the
	phosphorylated form of RB and the Rb-E2F complex) but they are not presented in the
	figure of the article and are deleted during the SEPI, we thus choose to not display
	them either.
   \textbf{B} True derivative of MyC (in blue) along with the ones
   computed by the CRN (in red), see main text for the exact definition of $D$.}
   \label{fig:switch}
\end{figure}

The model  \verb+biomodels 318+ of a bistable switch in the context of the restriction
point~\cite{yao2008bistable} displays a SEPI toward our derivative circuit.  This model,
presented in Fig.~\ref{fig:switch}\textbf{A}, study the Rb-E2F pathway as an example of
bistable switch where the presence of a (not modeled) growth factor activates the
MyC protein, starting the pathway until it reach the E2F factor that constitute the output
of the model. Yao \& al. show that once E2F reachs a threshold, its activation becomes
self sustained hence the notion of switch.

The SEPI given by Biocham is worth of interest as it does not merge any species and only
three reactions into one leaving all the other either untouched or deleted, thus
indicating that the pattern of the derivative is already well present. Morevoer, MyC is
mapped to the input and E2F to one part of the output, reinforcing our intuition that the
discovered SEPI is closed from the natural functionning of the CRN.

To conform this, we run the simulation as provided by the models and display the
derivative of the MyC protein against a scaled difference of the $D_+$ and $D_-$ species:
$D = a RB - b E2F$ where $a$ and $b$ are positive constant adjusted so that $D$ goes to
$0$ at final time and are of the same magnitude as $\frac{d MyC}{dt}$. (This gives $a=6.3,
b=0.063$.) Clearly, $D$ is a delayed and smoothed version of the input derivative exactly
as our derivative device would provide.

%
%
%

%
%
%
%
%

\section{Conclusion and perspectives}

We have presented a mathematical analysis of the core differentiation CRN
introduced by Whitby \& al.~\cite{WCKLT21ieee}. In particular, we have shown that what is
computed is an approximation of the left derivative given a small time in the past with a time
constant determined by the diffusion constant between the input and its internal
representation: $\epsilon = \frac{1}{\kd}$. Moreover, there is a delay $\tau$ due to the
computation time that can also be precisely estimated given the rate of activation and
degradation of the species encoding the derivative: $\tau = \frac{1}{k}$.
We have shown that such results can be used in some cases to design error-correcting terms
and obtain excellent implementations of functions of input signals using an approximation of their derivative on the fly.

From a synthetic biology perspective, the derivative CRN may be very relevant in the context of biosensor design,
when the test is not be about the presence of some molecular compounds~\cite{CAFRM18msb}
but on their variation. 
A derivative CRN is also needed to construct PID
controllers. The derivative control is known for damping the oscillations around the target of the
controller but delays are also known for producing such oscillations.
Being able to determine and quantify those delays and errors is thus important to optimize the design.
This device may also be used to approximate the derivative of an unknown
external input in the context of online cellular computing. Once again, delay may produce
nefarious artefacts that can easily be avoided when aware of the problem.

Furthermore, using the notion of SEPI to scan the biomodels database, we were able
to highlight a certain number of CRN models that contain the core differentiation CRN.
A high number of these models occur in models presenting oscillations.
We have shown on one such example, a circadian clock model, why it makes sense for an oscillator to
sense its own derivative, and to reproduce what a mathematician would produce in a more direct way for the
most basic oscillatory function: sine and cosine.

\subsection*{Acknowledgment}
This work benefited from ANR-20-CE48-0002 $\delta$ifference project grant.

\bibliographystyle{plain}
\bibliography{contraintes}

\section*{Appendix: computation of integration with a delay}

To prove that the drift of the output is a direct consequence of the delay, we first
compute the input and the approximate derivative for our choice of input:
\begin{equation}
	\begin{aligned}
		x(t) &= 1+\sin(t) \\
		x'(t-\tau) &= \cos(t-\tau) \\
		 &= \cos(t) + \tau \sin(t) + o(\tau^2)
	\end{aligned}
\end{equation}

Then we can compute the output up to the first order:
\begin{equation}
	\begin{aligned}
		y(t) &= \int 2 x(s) x'(s-\tau) ds \\
		 &= \int 2 \left(1+\sin(s)\right) \cos(s) ds + \int 2 \tau (\sin(s)+\sin^2(s)) ds  \\
		 &= \left(1+\sin(t)\right)^2 + 2 \tau \int \sin(s)+\sin^2(s) ds \\
		y(t) &\simeq \left(1+\sin(t)\right)^2 + \tau t
	\end{aligned}
\end{equation}

Then, to correct the observed drift, we propose to introduce a delay signal and use it in the
 computation to produce the output species $Y_+$ and $Y_-$, with the following CRN:
\begin{equation}
   \label{CRN:corrected}
	\begin{aligned}
		\Xext &\xrightarrow{\kd} \Xin, &\quad \Xin &\xrightarrow{\kd} \Xext, \\
		\Xin &\xrightarrow{k} \Xin + \Xd, &\quad \Xd &\xrightarrow{k} \emptyset, \\
		\Xext &\xrightarrow{\kd k} \Xext + D_+ &\quad D_+ &\xrightarrow{k} \emptyset \\ 
		\Xin &\xrightarrow{\kd k} \Xin + D_- &\quad D_- &\xrightarrow{k} \emptyset \\ 
		\Xd + D_+ &\xrightarrow{2} \Xd + D_+ + Y_+ &\quad
		\Xd + D_- &\xrightarrow{2} \Xd + D_- + Y_- \\
		D_+ + D_- &\xrightarrow{\text{fast}} \emptyset &\quad
		Y_+ + Y_- &\xrightarrow{\text{fast}} \emptyset \\
	\end{aligned}
\end{equation}

\end{document}